\documentclass{article}

\usepackage{PRIMEarxiv}

\usepackage[utf8]{inputenc} 
\usepackage[T1]{fontenc}    
\usepackage{hyperref}       
\usepackage{url}            
\usepackage{booktabs}       
\usepackage{nicefrac}       
\usepackage{microtype}      
\usepackage{lipsum}
\usepackage{fancyhdr}       

\usepackage{amsmath,amssymb,amsfonts}
\usepackage{algorithmic}
\usepackage{textcomp}

\usepackage{bm}
\makeatletter
\AtBeginDocument{\DeclareMathVersion{bold}
\SetSymbolFont{operators}{bold}{T1}{times}{b}{n}
\SetMathAlphabet{\mathrm}{bold}{T1}{times}{b}{n}
\SetMathAlphabet{\mathit}{bold}{T1}{times}{b}{it}
\SetMathAlphabet{\mathbf}{bold}{T1}{times}{b}{n}
\SetMathAlphabet{\mathtt}{bold}{OT1}{pcr}{b}{n}
\SetSymbolFont{symbols}{bold}{OMS}{cmsy}{b}{n}
\renewcommand\boldmath{\@nomath\boldmath\mathversion{bold}}}
\makeatother

\newtheorem{theorem}{Definition}
\newcommand{\digit}[1]{\vcenter{\hbox{\includegraphics[height=20pt]{img/#1}}}}
\DeclareMathOperator*{\argmin}{arg\,min}

\usepackage[colorinlistoftodos]{todonotes}

\title{A Synergistic Approach In Network Intrusion Detection By Neurosymbolic AI}

\author{
  Alice Bizzarri \footnotemark[2] \\
  University of Ferrara \\
  Ferrara, Italy \\
  \texttt{alice.bizzarri@unife.it} 
   \And
  Chung-En Yu \footnotemark[2] \\
  University of West Florida \\
  Pensacola, Florida, USA \\
  \texttt{cy31@students.uwf.edu} \\
   \And
  Brian Jalaian \\
  University of West Florida \\
  Pensacola, Florida, USA  \\
  \texttt{bjalaian@uwf.edu.org} 
   \And
  Fabrizio Riguzzi \\
  University of Ferrara \\
  Ferrara, Italy  \\
  \texttt{fabrizio.riguzzi@unife.it}
   \And
  Nathaniel D. Bastian \\
  United States Military Academy \\
  West Point, New York, USA \\
  \texttt{nathaniel.bastian@westpoint.edu}
}

\begin{document}
\maketitle

\footnotetext[1]{Both authors contributed equally to this research.}

\begin{abstract}
The prevailing approaches in Network Intrusion Detection Systems (NIDS) are often hampered by issues such as high resource consumption, significant computational demands, and poor interpretability. Furthermore, these systems generally struggle to identify novel, rapidly changing cyber threats. This paper delves into the potential of incorporating Neurosymbolic Artificial Intelligence (NSAI) into NIDS, combining deep learning's data-driven strengths with symbolic AI's logical reasoning to tackle the dynamic challenges in cybersecurity, which also includes detailed NSAI techniques introduction for cyber professionals to explore the potential strengths of NSAI in NIDS. The inclusion of NSAI in NIDS marks potential advancements in both the detection and interpretation of intricate network threats, benefiting from the robust pattern recognition of neural networks and the interpretive prowess of symbolic reasoning. By analyzing network traffic data types and machine learning architectures, we illustrate NSAI's distinctive capability to offer more profound insights into network behavior, thereby improving both detection performance and the adaptability of the system. This merging of technologies not only enhances the functionality of traditional NIDS but also sets the stage for future developments in building more resilient, interpretable, and dynamic defense mechanisms against advanced cyber threats. The continued progress in this area is poised to transform NIDS into a system that is both responsive to known threats and anticipatory of emerging, unseen ones.
\end{abstract}


\section{Introduction}

The emergence of artificial intelligence (AI) has revolutionized several sectors by enabling capabilities that were once beyond the reach of traditional computing technologies, as seen in neural network applications \cite{Kishor2022NeuroSymbolicAB}. Notably, AI has made a significant impact in the field of cybersecurity, especially in enhancing network intrusion detection systems (NIDS) \cite{drewek2020survey,dash2022threats}.

NIDS plays an essential role in safeguarding network integrity and privacy. However, with the escalating internet activity and corresponding increase in cyber threats, conventional NIDS are struggling to cope with the detection and mitigation of emerging attack vectors \cite{dash2022threats}. In response, AI technologies are increasingly being adopted as primary defense mechanisms in many organizations. Specifically, the fusion of AI with NIDS has proven to be effective in bolstering cybersecurity measures \cite{OzkanOkay2021IDSReview}.

In the specific area of NSAI, which merges rule-based AI with modern deep learning approaches, there is notable potential for progress in NIDS applications. For instance, NIDS models that combine deep learning with symbolic reasoning have shown impressive capabilities \cite{Sivatha2009neuro, Onchis2022NeuroSymbolicClassifierSecurity}. These models can refine their predictive accuracy through continuous feedback from system operators, reducing false positives. Additionally, Himmelhuber et al. \cite{Himmelhuber2022DetectExplainSymbolic} implemented methods that integrate knowledge graph learning and symbolic techniques to minimize false alarms and enhance result interpretability.

Various NSAI approaches exist, each distinguished by their method of blending symbolic logic with neural networks and their specific application contexts. Techniques like Logic Tensor Network (LTN) utilize tensors to represent logical formulas, integrating symbolic reasoning into neural frameworks \cite{badreddine2022logic}. Similarly, DeepProbLog and DeepStochLog merge probabilistic logic with deep learning \cite{manhaeve2021neural,winters2022deepstochlog}, and NeurASP incorporates neural networks within ASP logic programming \cite{yang2020neurasp}. Another innovative method, Semantic Loss, ensures the output from networks is semantically accurate by applying specific constraints \cite{xu2018semantic}.

This study aims to evaluate how NSAI can improve NIDS performance through a detailed AI analysis. We begin by assessing various datasets used in NIDS, highlighting the distinctions between network flow and packet-level data and comparing real-world versus synthetic datasets. We then explore relevant machine learning tasks for NIDS, such as anomaly detection and multiclass attack classification, and detecting unknown attacks through novelty and Out-of-Distribution (OOD) techniques. The conversation shifts to essential NIDS model characteristics, focusing on real-time processing capabilities, and the critical aspects of model interpretability and uncertainty qualification. Additionally, we review diverse model architectures from statistical models to advanced deep learning and sequential models, and the potential NSAI implementations in NIDS, discussing their strengths and weaknesses in addressing current NIDS challenges. Lastly, we bridge to introducing the details of NSAI, which further reveals the power of NSAI. This comprehensive analysis not only provides insights into the present landscape of AI-enhanced NIDS but also paves the way for future innovations incorporating NSAI.

\section{NIDS Datasets}

In the domain of NIDS, the choice and application of the right data types are crucial for refining detection mechanisms. Here, we delve into two essential elements of data utilization in AI-powered NIDS: evaluating the effectiveness of network flow information compared to packet-level data, and assessing the use of real-world data against synthetic data. Both data frameworks provide distinct benefits and present particular challenges that can notably affect the efficiency and performance of NIDS.

\subsection{Network Flow Information vs. Packet-level Data}

In AI-enhanced NIDS, the strategic use of two principal data types—network flow information and packet-level data—is crucial. Network flow data condenses network traffic into flows, providing a broader view of communication patterns which is useful for spotting potential intrusions like unusual traffic behaviors or spikes. This data type is particularly effective for monitoring large networks due to its reduced resource consumption \cite{Umer2017FlowBasedDetection}. On the other hand, packet-level data facilitates detailed packet inspection, examining each packet's headers and payloads. This allows for the identification of specific attack vectors, protocol anomalies, and complex threats that involve manipulation of payloads or encryption, offering a more nuanced and in-depth analysis of network traffic \cite{Farrukh2022PayloadByteTool}. Both data types are indispensable for a comprehensive intrusion detection capability, each providing distinct insights into network security.

A recent investigation \cite{Premkumar2023GRL} points out the shortcomings of conventional ML/DL methodologies in recognizing context-driven similarities between network flow information and packet-level data within NIDS. These traditional models generally specialize in analyzing either network flow or packet-level data. The research suggests that employing graph representation learning (GRL) can significantly boost NIDS efficiency by merging these two types of data. GRL techniques encode these relationships into a graph structure, capturing the complex and dynamic nature of network traffic and associated threats more effectively. The study proposes that graph-based data produced through GRL could be integrated into NSAI's symbolic reasoning frameworks, enriching them with extensive domain knowledge. This integration aims to enhance the predictive accuracy of NSAI and provide detailed, granular interpretability by utilizing packet-level data insights.

In this research, we argue that graph-based data, created through GRL methods, could be effectively incorporated into the symbolic reasoning system of NSAI, thereby enhancing its domain knowledge. Furthermore, an NSAI-enabled NIDS could harness both packet-level data and network flow insights, allowing for the detection of real-time threats by analyzing both payload and encrypted traffic. This integration of packet-level and network flow data in an NSAI-powered NIDS may significantly improve the effectiveness, resilience, and interpretability of these systems.

Ghadermazi et al. \cite{Ghadermazi2023RealTimeNIDImageBased} have introduced a technique that merges packet-level data with network flow information. This technique transforms sequential packets into two-dimensional images that are then processed using CNNs, facilitating the prompt detection and analysis of malicious actions. This method not only improves early detection over traditional flow-only methods but also handles the high computational demands and potential rise in false positives that come with processing large data volumes in real time. By utilizing a combination of packet and flow data, this approach maintains the temporal and spatial relationships among packets, thus improving the detection’s speed and accuracy.

The discussed case studies underline the vital importance of different network data types in NIDS methodologies. While packet-based detection excels at providing detailed insights into each packet, flow-based detection offers a broader view of network activity, crucial for spotting unusual traffic patterns. Furthermore, the combined method \cite{Premkumar2023GRL, Ghadermazi2023RealTimeNIDImageBased}, which leverages both types of data, presents a potent mechanism for real-time and precise threat detection. This hybrid strategy merges the detailed, immediate analysis of packets with the broader contextual understanding provided by flow data, setting the stage for further innovations in NIDS technology.

\subsection{Real-world vs. Synthetic Data}

In the realm of NIDS, the utilization of both real-world and synthetic data is crucial, each offering unique advantages and facing particular limitations. Real-world datasets, such as KDDCUP’99 \cite{KDDCup99}, NSL-KDD \cite{NSL-KDD}, CICIDS-2017 \cite{Sharafaldin2018CICIDS2017}, UNSW-NB15 \cite{Moustafa2015UNSWNB15}, and ACI-IOT-2023 \cite{Bastian2023ACIIOT2023}, are often derived from simulated real-environment conditions. These datasets are designed for the evaluation and benchmarking of NIDS solutions, providing labeled network traffic that includes a variety of attack types to enable thorough testing across different scenarios. However, the use of real-world data is frequently limited by issues related to privacy, data sensitivity, and other challenges, leading to an increased reliance on synthetic data.

Synthetic data, created using techniques such as generative adversarial networks (GANs), Markov models, and Bayesian networks, effectively replicates real-world traffic patterns and attack behaviors. This type of data is particularly useful for simulating rare or intricate attack scenarios, which enhances the robustness of NIDS \cite{Chale2022GenerateRealisticCyberData}. Additionally, techniques like those in Deep PackGen \cite{Hore2023DeepPackGen} allow for the generation of adversarial network packets, providing a stringent assessment of NIDS capabilities under realistic conditions. While synthetic data offers significant benefits due to its lack of privacy concerns and the ability for extensive sharing—thus promoting collaboration and improving reproducibility—it may not always match the variability and authenticity of real-world data. This mismatch can affect the performance of NIDS in genuine network settings. Moreover, generating high-quality synthetic data requires substantial computational resources and specialized knowledge, which can be a substantial hurdle for researchers with limited resources \cite{Chale2022GenerateRealisticCyberData}. Therefore, both types of data are essential, each contributing significantly to the enhancement and accuracy of NIDS algorithms in the field of cybersecurity.

\section{ML Tasks in NIDS}

NIDS is essential for protecting network infrastructures by quickly identifying and addressing various malicious cyber activities. The incorporation of Machine Learning (ML) and Deep Learning (DL) has significantly improved NIDS by providing advanced techniques for addressing key challenges such as anomaly detection, multiclass attack classification, and unknown attack detection. Each of these areas requires a sophisticated ML metrics to effectively evaluate the performance and reliability of NIDS in actual operational environments. This section discusses the essential performance metrics used in ML that help assess the detection capabilities of NIDS against various attack types.

In subsequent sections, we delve deeper into how these metrics are utilized in specific scenarios. This includes detecting anomalies, binary and multiclass classification for identifying different types of attacks, and specialized metrics for pinpointing unknown attacks. Each application plays a critical role in enhancing the robustness and effectiveness of intrusion detection systems, ensuring that they can respond accurately and efficiently to a broad spectrum of threats.

\subsection{General ML Metrics}

In the perspective of ML, we see detecting attacks as classification problems. And the performance metrics are critical in understanding how well of an NIDS detecting anomalies, classifying mutliclass attacks, or detecting unknown attacks. The following paragraphs introduce the common ML performance metrics used to evaluate NIDS effectiveness \cite{Thakkar2021SurveyIDSSystem}. For clarity and precision in documentation, the following abbreviations will be used: True Positives (TP), True Negatives (TN), False Positives (FP), and False Negatives (FN).

\textbf{Accuracy}: Accuracy measures the overall correctness of the model. In the context of NIDS, it is useful for binary classification, and high accuracy can indicate the system effectively distinguishes between normal and malicious traffic. However, in unbalanced datasets where attacks are rare compared to normal traffic, accuracy may not be the most reliable metric, as a model could wrongly predict most samples as the majority class and still achieve high accuracy. The accuracy can be calculated by:

\begin{equation}
    \text { Accuracy }=\frac{T P+T N}{T P+T N+F P+F N}
    \label{eq:accuracy}
\end{equation}

\textbf{Precision}: Precision measures the accuracy of positive predictions, high precision indicates that the model is reliable when it predicts in binary classification, which is formulated as:

\begin{equation}
    \text { Precision }=\frac{T P}{T P+F P}
    \label{eq:precision}
\end{equation}

\textbf{Recall}: Recall, or true positive rate (TPR), measures the model's ability to capture all relevant attacks, calculated as Eq. \ref{eq:recall}. High recall is essential in NIDS to ensure that few anomalies are missed, and is also essential for for unknown attacks where no previously identified signature exists.

\begin{equation}
    \text { Recall }=\frac{T P}{T P+F N}
    \label{eq:recall}
\end{equation}

\textbf{True Negative Rate (TNR)}: TNR, or specificity, measures the proportion of actual negatives that are correctly identified (e.g., the proportion of normal activities correctly identified as such), formulated as Eq. \ref{eq:tnr}. High TNR means the system effectively identifies normal behavior, reducing the likelihood of false alarms.

\begin{equation}
    \text { True Negative Rate }=\text { Specificity }=\frac{T N}{T N+F P}
    \label{eq:tnr}
\end{equation}

\textbf{Confusion Matrix}: Confusion Matrix lays out the TP, FP, TN, and FN, providing a clear picture of the model’s performance across the different classes. It helps in visualizing the performance of a classification algorithm. It is useful for anomaly detection, binary and multiclass attacks classification.

\textbf{F1 Score}: F1 Score is the harmonic mean of precision and recall, providing a balance between the two when the class distribution is uneven. It's particularly useful in NIDS when dealing with anomaly detection and unknown attacks, where failing to detect even a few instances can be catastrophic. The F1 Score helps mitigate the bias toward the majority class that can affect the accuracy metric. It can be formulated as Eq. \ref{eq:f1}.

\begin{equation}
    \text { F1 Score }=2 * \frac{\text { precision } \times \text { recall }}{\text { precision }+ \text { recall }}
    \label{eq:f1}
\end{equation}

\textbf{Macro F1 Score}: Macro F1 Score computes the F1 Score independently for each class but does not use class imbalance as a weight, giving equal weight to all classes. This is useful in multiclass attacks classification scenarios within NIDS, ensuring that rare attack types are considered as significant as common ones, which helps in training models to recognize less frequent but potentially more harmful attacks. Macro F1 Score can be formulated as \eqref{eq:macrof1}.

\begin{equation}
    \text { Macro F1 Score}=\frac{1}{N} \sum_{i=1}^N \text { F1 Score}_i
    \label{eq:macrof1}
\end{equation}

\textbf{Area Under the ROC Curve (AUROC curve)}: AUROC curve evaluates the model's ability to discriminate between classes at various threshold settings. A higher AUROC value indicates better model performance and a greater likelihood of correctly distinguishing between normal and malicious activities. This metric is particularly useful for anomaly detection, binary attacks classification and unknown attacks detection in NIDS, allowing administrators to adjust the sensitivity of their intrusion detection systems based on the risk level of missing an attack versus raising false alarms. 

\textbf{Matthews Correlation Coefficient (MCC)}: MCC is a more informative metric than F1 score and accuracy as it takes into account true and false positives and negatives and is generally regarded as a balanced measure which can be used even if the classes are of very different sizes. This can be used for binary attacks classification. It is formulated as Eq. \ref{eq:mcc}.

\begin{equation}
    \text{MCC}=\frac{T P \times T N-F P \times F N}{\sqrt{(T P+F P)(T P+F N)(T N+F P)(T N+F N)}}
    \label{eq:mcc}
\end{equation}

\subsection{Anomaly Detection}


In NIDS, the goal of anomaly detection is to spot irregularities that diverge from normal network behavior, potentially signaling security threats. Advanced AI methodologies are increasingly pivotal in the realm of anomaly detection. One notable approach is detailed by Hore et al. \cite{Hore2023AEAnomalyDetection}, who utilize autoencoders—neural networks geared towards unsupervised learning—to train solely on benign traffic. This training helps establish a baseline of normal activities, enabling the detection of abnormal behaviors indicative of intrusions. This method leverages packet-based data to facilitate the real-time analysis necessary for NIDS, ensuring swift responses to emergent threats.

Additionally, Bierbrauer et al. \cite{Bierbrauer2021AnomalyDetectionAdversarial} explore anomaly detection in adversarial settings, where attackers might manipulate detection systems. Their study integrates unsupervised learning with graph-based techniques and an innovative supervised stacking ensemble method. At the secondary level, this ensemble employs classifiers such as Naive Bayes and Decision Trees, achieving an F1 Score above 0.97 for detecting malicious activities, thus demonstrating robust classification efficacy. This approach also emphasizes the utilization of packet-level data, crucial for real-time threat detection and processing in dynamic network environments.

The surge in intrusion attacks has notably increased the reliance on AI-driven NIDS. However, machine learning and deep learning methods deployed in these systems often face difficulties due to high rates of false positive (FP) alerts and a lack of interpretability. These issues can significantly add to the workload of cybersecurity professionals who manage these alerts. In response, a notable study by Himmelhuber et al. \cite{Himmelhuber2022DetectExplainSymbolic} introduced an innovative approach using knowledge graph learning techniques coupled with symbolic methods. Specifically, they employed Graph Neural Networks (GNNs) and the GNNExplainer tool \cite{Ying2019Gnnexplainer} to not only detect anomalies but also provide explanations for their detections.

The explanations generated by GNNExplainer were then used to create a filter aimed at reducing the frequency of FP alerts. The results from this study proved significant, as they not only decreased the number of unnecessary alerts but also improved the interpretability of the detections. This advancement provides substantial operational benefits, enhancing the efficiency and effectiveness of network security management in NIDS environments.

\subsection{Multiclass Attacks Classification}


Detecting multiclass attacks in NIDS involves categorizing network traffic into various attack classes to accurately identify and respond to different cyber threats. This process is treated as a multiclass classification problem in machine learning.  Ahmad et al. \cite{Ahmad2020NIDSystemML} conducted a thorough analysis of various ML/DL techniques for classifying multiclass attacks using well-known datasets like KDD Cup '99 \cite{KDDCup99} and NSL-KDD \cite{NSL-KDD}. Their study explored a variety of ML algorithms including Decision Trees (DT), K-Nearest Neighbors (k-NN), and K-Means Clustering, along with DL methods such as Recurrent Neural Networks (RNNs), Convolutional Neural Networks (CNNs), and Autoencoders (AEs). They pointed out the significant issue of imbalance among minority attack classes, which substantially affects the detection rates. The study also highlighted the necessity of employing intelligent feature selection techniques to mitigate the challenges of deploying ML and DL models, which are often resource-intensive and time-consuming.

In another study, Bedi et al. \cite{Bedi2021ISiamIDS} focus on the Improved Siam-IDS (I-SiamIDS) system, which utilizes a two-layer ensemble architecture to address class imbalances effectively. The second layer of this system employs a multiclass eXtreme Gradient Boosting (m-XGBoost) classifier designed to categorize various detected attacks into distinct classes like Denial of Service (DoS), Probe, Remote to Local (R2L), and User to Root (U2R). This approach significantly enhances performance metrics such as Accuracy, Recall, Precision, and F1 Score for each attack category, showcasing its ability to manage not just the majority classes but also the underrepresented minority classes effectively.

Additionally, Maseer et al. \cite{Maseer2021BenchmarkMLNIDS} conducted an extensive evaluation of various ML algorithms for detecting and classifying multiple network attack types using the simulated real-world CICIDS2017 dataset. The study covered both supervised and unsupervised models, including DT, k-NN, Naive Bayes (NB), Random Forest (RF), Support Vector Machines (SVM), and CNNs, assessing their performance across metrics like accuracy, precision, recall, and F1 score. Notably, models like k-NN, DT, and NB were emphasized for their excellent detection capabilities and computational efficiency, making them particularly suitable for multiclass classification tasks. However, the study also revealed the challenges posed by class imbalance, which can skew performance metrics in favor of more frequently occurring classes. Moreover, the complexity of tuning and configuring these models to achieve optimal performance was noted as a practical barrier, especially for complex models like CNNs that demand considerable computational resources.

Despite concerted efforts to tackle imbalanced attack classes in NIDS, challenges persist due to the resource-intensive and time-consuming nature of traditional ML/DL models, as discussed in studies such as that by Bedi et al. \cite{Bedi2021ISiamIDS}. To address these concerns, a novel approach in the realm of NSAI employs the Logic Tensor Network (LTN), initially proposed by Serafini et al. \cite{Serafini2016LTN}. LTNs blend tensor network-based learning with logical reasoning, creating a sophisticated framework particularly well-suited for complex relational and classification tasks, especially within multiclass learning environments.

LTNs leverage logical and domain-specific rules to articulate relationships and class implications within data, enhancing the learning process by fostering more balanced class representations, even in scenarios where certain classes are notably underrepresented. This ability makes LTNs adept at tackling the pervasive issue of imbalanced datasets in NIDS. Moreover, LTNs facilitate a more systematic approach to model configuration. Adjustments to the logical layer can modify the learning process, circumventing the extensive parameter tuning typically required in deep learning models. Overall, LTNs offer a promising strategy to both mitigate the effects of imbalanced datasets and simplify the configuration of machine learning models. This advancement enhances the performance and utility of classification systems in NIDS by incorporating domain-specific rules that aid in the detection of various types of network attacks, making LTNs a valuable asset in improving cybersecurity measures.

\subsection{Unknown Attacks Detection}

Unknown attacks in NIDS refer to previously unknown attack scenarios that are absent from the training datasets of ML models. These attacks are particularly challenging because they lack existing signatures or recognizable patterns, rendering them difficult to detect. Within the realm of ML, unknown attack detection is approached as a binary classification task. Unknown attacks are unique in that they have no prior known signatures or profiles since they have not yet been identified in real-world settings. 


In the context of NIDS, it’s crucial to distinguish between novelty detection and Out-of-Distribution (OOD) detection. Novelty detection aims to identify instances that significantly diverge from the known data profiles established during the training phase. This method focuses on spotting anomalies that stray from expected behaviors, thereby enabling NIDS to adjust to new threats and improving their capacity to recognize attacks that have never been seen before. Conversely, OOD detection targets the identification of instances that fall outside any recognized class or distribution within the dataset. This approach is critical for flagging activities that do not match any known attack patterns, thus aiding NIDS in detecting potentially suspicious actions that might signify unknown attacks.

    \subsubsection{Novelty Detection}
    A novel technique involving the Cyber Creative Generative Adversarial Network (CCGAN) has been developed to enhance novelty detection in NIDS by generating synthetic network packets \cite{Pavlik2023CyberCreativeGAN}. CCGAN uses generative modeling to train on current NIDS datasets, producing synthetic malicious packet payloads that differ from established attack profiles. This method allows NIDS to stay ahead of evolving cyber threats by identifying new anomalies that traditional detection systems might miss, thus improving the system’s capability to detect emerging cyber threats. 
    
    However, there are critical concerns with this approach, primarily concerning the authenticity of the synthetic data. It's essential that this data closely mirrors real network traffic to validate the model's effectiveness in actual scenarios. Another significant issue is the model’s ability to generalize to entirely unknown attack types not included in the training data. The success of CCGAN also heavily relies on the quality and breadth of the datasets it trains on; outdated or narrow datasets could hinder the utility of the synthetic packets it generates. Additionally, the computational requirements and scalability of GANs might pose logistical challenges, especially in environments where resources are limited or where real-time processing is crucial, presenting substantial operational challenges.

    Incorporating the Cyber Creative Generative Adversarial Network (CCGAN) for discovering novel patterns, recent advancements also highlight the effectiveness of survival analysis in identifying critical attributes that detect novel network traffic types. The study by Bradley \cite{Bradley2023NoveltyDetectionNetworkTraffic} utilized Cox proportional hazards models and Kaplan-Meier estimates to forecast the detection likelihood of new network flows by classifiers after new types of attacks are introduced. Key factors such as PSH Flag Count, ACK Flag Count, URG Flag Count, and Down/Up Ratio were determined to significantly affect novelty detection in models like Random Forest, Bayesian Ridge, and Linear Support Vector Regression. Incorporating these survival analysis techniques allows NIDS to better recognize and adapt to emerging cyber threats, enhancing their defensive capabilities.
    
    \subsubsection{Out-of-distribution (OOD) Detection}
    An Out-of-Distribution (OOD) detection study by Hore et al. \cite{Hore2023SequentialDLNIDS} presents a framework that employs a sequential deep neural network (DNN) model to analyze network traffic data, effectively identifying OOD attack patterns. The sequential nature of the model is pivotal as it captures temporal dependencies and anomalies in network behavior over time, essential for detecting sophisticated cyber attacks that may elude traditional NIDS. The model has shown considerable success, achieving an average accuracy of 98.5\% in detecting OOD attacks.
    
    However, there are significant challenges associated with this approach, primarily due to the computational complexity of utilizing multiple sequential DNN architectures. This complexity can lead to heightened computational demands, potentially impeding the framework's scalability and efficiency, especially in environments where real-time processing is necessary. Additionally, the model's reliance on one-shot learning, although innovative, introduces further complications. One-shot learning necessitates that the model effectively learn from a very limited amount of data, which can cause overfitting or underfitting if not carefully handled. Consequently, the framework's performance is highly dependent on the quality and representativeness of the training samples utilized during the one-shot learning phase, making it particularly sensitive to the adequacy of its training data.

    Farrukh et al. \cite{Farrukh2023AISNIDS} have developed a sophisticated framework for NIDS that combines packet-level analysis with autonomous learning features. An Intelligent and Self-Sustaining Network Intrusion Detection System (AIS-NIDS) utilizes a complex immune network model, drawing inspiration from biological immune systems, to autonomously differentiate between normal and novel attack vectors while continuously evolving to recognize new threats. This approach integrates deep learning for detailed packet inspection, enhancing the precision of threat detection. Additionally, AIS-NIDS is equipped with an incremental learning feature that allows the system to update its database with new patterns autonomously, minimizing the need for frequent retraining and thus reducing operational downtime. Extensive real-world testing has validated the effectiveness of AIS-NIDS in identifying and adapting to new categories of attacks, thereby bolstering the security of network infrastructures.
    
    On a related note, recent advancements in tackling the detection of OOD attacks in NIDS emphasize the relevance of Open Set Recognition (OSR) techniques. Baye et al. \cite{Baye2023PerformanceAnalysisDL} explored several deep learning-based OSR algorithms, assessing their effectiveness in recognizing unknown attack types within network systems. Their research introduced the OpenSetPerf framework, a tool for evaluating how different hyperparameters affect the performance of OSR algorithms using a standard NIDS dataset. The results indicate that energy-based OOD detection methods and OpenMax algorithms significantly outperform traditional Softmax classifiers in identifying unseen threats. These innovations in OSR technology enable NIDS to better adjust to the evolving dynamics of network traffic, enhancing their capacity to detect and respond to new cyber threats effectively.
    
\indent
NSAI provides innovative approaches that significantly enhance the detection of unknown attacks by merging neural representations with logical rules. These methods leverage the computational power and adaptability of deep learning within neural networks alongside the structured, rule-based reasoning of symbolic AI. This integration aims to improve not only the accuracy but also the interpretability, efficiency, and scalability of the detection models.

A basic example of how symbolic frameworks contribute to NSAI is through the use of simple decision trees, which rely on propositional logic. These trees serve as a foundational level-0 application within NSAI, employing straightforward logical structures to classify data, thus boosting the model's interpretability. Progressing to a level-1 NSAI involves enhancing these capabilities by incorporating the ability to identify and learn relationships between different entities within the input data, thereby potentially increasing the effectiveness and precision in detecting unknown attacks.

Furthermore, NSAI enhances detection rates by integrating logical rules into a loss function, which strategically utilizes the strengths of both its neural and symbolic components. The inherent efficiency of the tree structure also offers potential advantages over more computationally demanding DL models, presenting a more streamlined alternative that can be particularly valuable in scenarios requiring rapid processing and response. This methodological framework not only strengthens the detection capabilities but also makes the overall system more robust and adaptable to new and emerging threats.

\section{NIDS Model Properties}

\subsection{Real-time NIDS}

Recent innovations in NIDS have greatly enhanced the capacity for real-time threat detection, introducing several groundbreaking methods that improve both the speed and accuracy of detecting network attacks.

One such advancement is introduced by Ghadermazi et al. \cite{Ghadermazi2023RealTimeNIDImageBased}, who propose a technique where sequential network packets are converted into two-dimensional images for analysis using Convolutional Neural Networks (CNNs). This approach allows for the detection of attacks as they occur, significantly reducing response times compared to traditional flow-based methods. However, this real-time processing requires substantial computational resources and sophisticated algorithms, which could increase the complexity and operational costs of the system. Additionally, maintaining high reliability to minimize false positives and negatives is crucial for the effectiveness of the operation.

Another innovative real-time detection method is presented in \cite{Chen2023NIDSviaDT}, which utilizes decision transformers to analyze packet sequences for prompt and informed decision-making. The decision transformer model is particularly adept at processing sequential data, thus enhancing the timeliness and accuracy of threat detection. Nonetheless, this technique demands significant computational power due to the complexities involved in real-time data processing and also depends heavily on the quality and structure of the input data.

A further advancement in real-time detection is described in \cite{Chen2023RIDE}, which details a software-hardware co-design framework integrating recurrent autoencoders and decision trees within a memristor-based architecture. This approach leverages memristor technology to enable instant processing of packet-level data, crucial for timely detection and response to network threats. While this method significantly boosts detection speeds and accuracy, it also faces challenges such as high computational demands and the nascent nature of memristor technology, which might pose novel design and maintenance challenges.

These examples illustrate diverse approaches to significantly reducing detection latency and enhancing accuracy, yet they also underscore the inherent challenges associated with deploying advanced computational technologies in real-time settings. The adoption of these cutting-edge technologies marks a critical advancement in the capabilities of intrusion detection systems, addressing the increasingly complex and swift nature of cyber attacks.

\subsection{Interpretability}

Interpretability is a crucial feature of AI-driven NIDS to ensure transparency, trustworthiness, and effective decision-making. Traditional ML models in NIDS, like Decision Trees (DT) and Support Vector Machines (SVM), offer a degree of interpretability by exposing the decision rules or support vectors that underlie classifications \cite{Miller2019AIExplanation}. These models, however, often struggle to detect modern, complex cyber attacks, necessitating more advanced solutions. Nevertheless, the interpretability of more sophisticated ML models, such as Artificial Neural Networks (ANN), can be limited, posing challenges in understanding the bases of their predictions. While DL models, including CNNs and RNNs, significantly improve detection capabilities, their opaque, black-box nature can impede interpretability in NIDS \cite{GhoshRoy2022ExplainableAI}.

On the other hand, NSAI seeks to integrate the pattern recognition capabilities of neural networks with the logical reasoning of symbolic AI, aiming to enhance both detection accuracy and interpretability in NIDS. However, achieving improvements in performance and interpretability simultaneously through NSAI can be challenging, as discussed in the work by Jalaian et al. \cite{Jalaian2023NSAICybersecurity}. While the symbolic AI components improve the system’s interpretability by providing structured and explicit reasoning in the decision-making process, they can also restrict the model's capacity to recognize new, complex patterns that neural networks handle more adeptly. This limitation arises because symbolic systems rely on predefined knowledge and rules, which may not adapt swiftly to novel attack vectors not previously encoded in the system. Conversely, while neural networks are proficient at detecting such novel patterns through data-driven learning, their "black box" nature obscures the decision-making process, thus diminishing the system’s transparency.

Despite these obstacles, integrating NSAI into NIDS offers a substantial opportunity to bolster cybersecurity defenses. This fusion not only promises enhanced detection of complex and evolving threats but also aims to increase trust in automated security systems by making their decisions more interpretable and justifiable. For cybersecurity professionals, this translates into an improved ability to comprehend and respond to the system’s alerts, potentially leading to quicker and more effective mitigation strategies.

\subsection{Uncertainty Qualification}

A study \cite{Wong2023UncertaintyQuantifiedDLNID} discusses the role of uncertainty quantification (UQ) within NIDS, especially its effectiveness in countering unknown attacks. UQ in this context involves integrating confidence measures into the predictions of ML models to identify out-of-distribution (OOD) attacks, which are typically not represented in the training data. This is essential in the field of cybersecurity, where the nature of threats evolves swiftly, often departing from known patterns. By utilizing Bayesian deep learning techniques and Hamiltonian Monte Carlo (HMC) approaches, the study introduces a way to quantify uncertainty. This approach allows models to evaluate the probability distributions associated with their predictions and allocate appropriate confidence levels. A prediction marked by high uncertainty might indicate an OOD attack, suggesting a new threat type that wasn't encountered during the training phase. This system's utility lies in its ability to not just identify potential attacks but also to provide a probabilistic certainty measure, thereby informing more cautious decision-making in situations of high uncertainty.

A study referenced in \cite{MatejekSafeguardingNID} underscores the pivotal role of UQ in bolstering the defenses of NIDS against unforeseen zero-day attacks and concept drift. This technique leverages generative models to map out class-specific internal feature representations within deep neural networks (DNNs). It introduces a protective mechanism that assesses the uncertainty of the classifier’s judgments, effectively pinpointing new and anomalous inputs with an impressive AUROC score exceeding 0.97. This mechanism employs normalizing flows to approximate the features' probability density function, thus identifying OOD events, typically indicative of novel attack vectors. This strategy not only aids in anomaly detection but also quantifies the confidence level of each detection, significantly enhancing the decision-making capability in network environments that continuously encounter new threats.

On the other hand, implementing UQ comes with its own set of challenges. While it facilitates the identification of unknown threats by highlighting uncertain predictions, it also increases the complexity of the model's architecture and adds to the computational load. Bayesian networks, particularly, demand significant computational resources as they involve sampling from intricate posterior distributions of model parameters—a process not easily scalable in real-time operational settings. Moreover, managing the trade-off between detection accuracy and uncertainty management poses a considerable hurdle. A model may achieve high accuracy but might struggle with uncertainty management, possibly resulting in overconfidence in its predictions. Conversely, prioritizing uncertainty can cause a model to become overly cautious, potentially increasing the rate of false positives.

\subsection{Model Architecture}

The structure and design of detection models are crucial for the effective and efficient identification and mitigation of cyber threats. With the escalating sophistication of cyber attacks, the importance of developing advanced model architectures capable of adapting to and countering new threats in real-time is increasingly critical. This discussion focuses on the variety of model architectures utilized in NIDS, examining their functionalities, benefits, and the challenges they face. Through a comprehensive review of these varied approaches, the aim is to demonstrate how distinct model architectures enhance the robustness and flexibility of NIDS, providing valuable perspectives on the practical applications of intrusion detection technologies.

    \subsubsection{Statistical Models}
    
    In NIDS, a statistical model functions by observing typical transactions to formulate a baseline profile of legitimate user behavior. Any deviation from this norm that surpasses a predetermined threshold is flagged as a potential security threat. These models are notably effective because they can identify novel attacks without relying on prior knowledge of attack patterns, require minimal maintenance since they do not need regular updates, and excel in detecting DoS and DDoS attacks. However, there are inherent challenges such as the significant time investment needed to develop a normal user profile, the dynamic nature of this profile which can evolve, and the requirement for statistical methods to be both accurate and robust to ensure ongoing system dependability \cite{OzkanOkay2021IDSReview}.
    
    \indent
    Regarding specific statistical measures, standard deviation plays a vital role in NIDS by helping to identify anomalies within network traffic data. It aids in calculating a profile's byte frequency distribution during an NIDS's training phase, setting a baseline for normal patterns and spotting deviations that might signify intrusions. The use of standard deviation in defining normal behavior and identifying outliers is well documented, including in works by \cite{Wang2004AnomalousPayloadNID}. Additionally, the integration of standard deviation in the feature extraction process can distinguish between normal and anomalous network behaviors effectively \cite{Ahmad2020NIDSystemML}.
    
    \indent
    The Z-Score is another statistical method used in NIDS that normalizes and standardizes network data, aiding in the identification of outliers and potential indicators of security threats \cite{Wang2022NovelMethodNID}. Through the application of Z-Score normalization, NIDS are equipped to more accurately identify and monitor deviations from established norms of behavior, thereby enhancing the detection capabilities for suspicious activities. This methodology is detailed in research such as \cite{Wang2022NovelMethodNID}, where the authors examine network data characteristics and implement Z-Score in conjunction with TF-IDF to standardize the data. This standardization aids in constructing a fuzzy triadic background with quadratic features. The authors showcase how combining Z-Score with other preprocessing techniques optimizes network data for more effective intrusion detection.

    
    
    \indent
    Multivariate correlation analysis (MCA) is also used within NIDS to detect DoS attacks by analyzing correlations among multiple variables in network traffic, thus identifying unusual patterns that might indicate attacks. This approach not only improves the accuracy of real-time detection but also helps reduce false positives by considering the complex interdependencies among network features. However, MCA requires comprehensive baseline data to define normal patterns and may suffer from high false positive rates if legitimate network behavior changes, along with being computationally intensive in high-traffic scenarios \cite{Tan2013MCA}.

    \subsubsection{Classic ML models}
    
    Classic Machine learning (ML) model has become an integral component of NIDS, providing advanced techniques for detecting malicious activities in network traffic. Classic ML methods in NIDS can be broadly classified into supervised and unsupervised learning approaches, each with distinct capabilities and challenges when applied to cybersecurity \cite{Thakkar2021SurveyIDSSystem}.
    
    Supervised learning algorithms require labeled data to train models that can classify or predict malicious behavior based on historical patterns. Advantages of supervised learning include high accuracy and the ability to adapt to new threats quickly if properly trained. Disadvantages include dependency on large, well-labeled datasets, which can be expensive and time-consuming to prepare. It also struggles with unknown attacks that do not match any previous patterns. Examples of supervised learning in NIDS include:
    \begin{enumerate}
        \item \textbf{Support Vector Machines (SVM)}: Used for classifying network traffic as normal or malicious based on a hyperplane that separates different classes.
        \item \textbf{Decision Trees (DT)}: These models use branching methods to make decisions, ideal for rules-based classification.
        \item \textbf{Random Forests (RF)}: An ensemble of decision trees that improves classification accuracy and overcomes overfitting issues present in individual decision trees.
    \end{enumerate}
    
    Unsupervised learning algorithms do not require labeled data. They identify patterns based on the inherent structure of the data. Advantages include the ability to detect novel attacks by identifying outliers or anomalies in traffic behaviors. Disadvantages encompass challenges in distinguishing between benign and malicious anomalies without prior labels, potentially leading to higher false positive rates. Examples of unsupervised learning in NIDS include:
    \begin{enumerate}
        \item \textbf{k-Means Clustering}: Used to segment network traffic into clusters based on similarity, which can help identify unusual patterns.
        \item \textbf{Principal Component Analysis (PCA)}: A technique used to reduce the dimensionality of data, enhancing the detection process by focusing on the most significant features.
    \end{enumerate}
    
    In summary, ML provides powerful tools for enhancing NIDS, with supervised methods excelling in environments with rich, labeled datasets and unsupervised approaches offering flexibility in detecting new and emerging threats. However, both categories face challenges such as the need for extensive data preparation and the risk of misclassifying new or complex attack vectors. Effective implementation of ML in NIDS requires a careful balance between model complexity, data availability, and the operational demands of real-time network security monitoring.
    
    \subsubsection{DL models}
    
    The incorporation of deep learning (DL) models into NIDS offers transformative potential to identify and counteract sophisticated cyber threats efficiently. Deep learning leverages complex neural network architectures that can learn and make inferences from large volumes of data, enabling them to detect patterns and anomalies that traditional methods might miss \cite{Thakkar2021SurveyIDSSystem}. We list three popular DL methods in NIDS: 
    \begin{enumerate}
        \item \textbf{Convolutional Neural Networks (CNNs)}: CNNs are frequently utilized in NIDS to autonomously identify hierarchical features within network traffic, aiding in the categorization of traffic as either benign or malicious based on recognized patterns. These networks excel in extracting features directly from raw inputs like traffic flows or packet details, eliminating the need for manual feature engineering. This capability is especially beneficial for processing image-based or sequential data where understanding spatial relationships is essential. However, the deployment of CNNs involves significant computational demands for both training and inferencing phases. Additionally, their efficacy is largely contingent on the availability of extensive labeled datasets. A notable limitation of CNNs is their reduced interpretability, which poses challenges in security contexts where explaining the basis of a detection decision is imperative.
        \item \textbf{Long Short-Term Memory Networks (LSTMs)}: LSTMs are adept at monitoring continuous streams of network data, detecting anomalies indicative of potential security threats through deviations from established patterns. Their proficiency in managing sequential data makes them ideal for analyzing time-series network traffic. LSTMs are particularly effective at recognizing long-term dependencies within data sequences, an attribute that is crucial for identifying sophisticated, multi-stage cyber-attacks. However, training LSTMs can be challenging; they are susceptible to overfitting and require substantial computational resources, which may hinder their deployment in real-time systems. Similar to CNNs, LSTMs also suffer from limited interpretability, a significant drawback in security applications where understanding the reasoning behind decisions is crucial.
        \item \textbf{Autoencoders (AEs)} AEs are utilized in NIDS to replicate normal network traffic patterns and identify deviations that may suggest potential intrusions. In unsupervised learning scenarios, AEs prove effective for anomaly detection. They are trained to reconstruct what is considered normal behavior, and anomalies are flagged when reconstructions significantly differ from these expected outputs, making AEs particularly adept at spotting unknown attacks. Nonetheless, calibrating autoencoders to accurately differentiate between benign and malicious anomalies presents a challenge and can lead to increased rates of false positives. Additionally, optimizing their performance necessitates meticulous tuning of the loss functions and the dimensions of the latent space.
    \end{enumerate}
    
    DL models equip NIDS with the capability to automatically and efficiently process and analyze extensive network data, enabling learning and detection of both familiar and novel threats. Although these models face challenges like substantial computational requirements, the necessity for large datasets, and difficulties in model interpretability, DL techniques are still advancing the capabilities of NIDS. Their aptitude for learning directly from data and recognizing intricate patterns offers a substantial improvement over traditional ML methods, positioning them as an essential element of contemporary cybersecurity defenses \cite{Thakkar2021SurveyIDSSystem}.
    
    \subsubsection{Sequential Models}
    
    In the dynamic field of NIDS, the demand for models that can promptly adapt to emerging and complex cyber threats is more urgent than ever. Traditional ML and DL methods often rely on static models trained on historical data, which may become obsolete as new threats arise. This results in a continuous need for updates and retraining, consuming substantial resources. Recent advancements have seen a shift towards more sophisticated, sequential models that enable real-time adaptability and decision-making capabilities.
    
    One innovative example of such a sequential model is Deep PackGen, introduced by Hore et al. \cite{Hore2023DeepPackGen}. This model utilizes Deep Reinforcement Learning (DRL) to generate adversarial network packets, focusing on sequential decision-making. The AI agent in this model learns to strategically alter packets to disguise malicious activities as benign, while maintaining their detrimental intent. Deep PackGen's dynamic adaptability sets it apart from traditional ML and DL approaches, potentially reducing the frequency of model updates and retraining. However, its practical application may be hindered by the dual challenges of maintaining packet functionality during modifications and the high computational demands of training DRL models.
    
    Another advanced sequential model is the decision transformer, utilized by Chen et al. \cite{Chen2023NIDSviaDT}. This model is built around the concept of sequence modeling for real-time decision-making, treating intrusion detection as a series of decision points rather than static classifications or predictions. Utilizing causally masked transformers, it conditions on historical sequences of network packets and previous detection decisions to forecast future actions, thereby enhancing real-time decision-making based on observed network activity patterns. Unlike static ML/DL models, the decision transformer can continually learn from ongoing network activities, improving its accuracy and responsiveness to new threats. However, the need for large volumes of sequential data and the computational intensity required for real-time processing are potential drawbacks in operational environments.
    
    Both Deep PackGen and the decision transformer signify major progress in NIDS by emphasizing dynamic, real-time approaches to manage the constantly evolving landscape of network threats. Nevertheless, the complexity and ongoing maintenance these models require due to their computational demands represent significant challenges for their practical deployment.
    
    \subsubsection{NSAI Models}

    The fusion of NSAI in cybersecurity realms brings substantial advancements, particularly in NIDS. NSAI, a blend of neural networks and symbolic AI, taps into the strengths of both technologies: neural networks' proficiency in handling large datasets and symbolic AI's robust reasoning capabilities. This synergy results in heightened accuracy and interpretability in detecting cyber threats, making NSAI-driven NIDS more adept at navigating the complexities of modern cyberattacks compared to conventional ML and DL approaches \cite{Jalaian2023NSAICybersecurity, Piplai2023KnowledgeEnhancedNeurosymbolicAI}.

    Techniques such as Logic Tensor Networks (LTNs) and Neural Logic Machines (NLMs) exemplify NSAI's potential. LTNs weave symbolic logic within neural frameworks, integrating logical operators directly into computations and enhancing relational data reasoning. NLMs, on the other hand, convert logical predicates into neural-friendly tensors, merging deep learning's pattern recognition with logical analysis. These methods significantly bolster NIDS's capabilities in preempting and reacting to unknown attacks.

    An exemplary use of NSAI in cybersecurity is found in the development of a neurosymbolic classifier that monitors security alerts in network traffic \cite{Onchis2022NeuroSymbolicClassifierSecurity}. This classifier utilizes Logic Tensor Networks (LTN) for the multiclass classification of network flow data using the KDD-Cup’99 dataset. It introduces a methodological enhancement by incorporating real logic to formulate complex queries, improving data interpretation and classification precision. A distinctive feature of this method is the implementation of a set of axioms that delineate relationships among various network connection types, which assists in accurately classifying normal and attack traffic. The classifier's effectiveness is measured by several criteria, including Knowledge Base satisfiability, training and testing accuracy, and the satisfiability of specific logical formulas that delineate the interconnections among various network activities. The findings reveal considerable detection accuracy improvements via neurosymbolic integration, enabling superior generalization across diverse network intrusion types. Achieving high satisfiability and accuracy, the classifier showcases the potential of merging symbolic AI with DL to forge more resilient and interpretable systems for NIDS. This strategy not only bolsters the classifier’s prowess in detecting familiar network attack types but also improves its adaptability to unknown attacks.

    However, the advancement of NSAI in NIDS faces several hurdles. A principal challenge involves the integration of symbolic and neural elements to enable fluid interaction and learning. This challenge extends to developing techniques that translate between symbolic and neural representations and integrating the strengths of both paradigms into a unified model, potentially requiring cybersecurity expertise for effective deployment and upkeep. Additionally, the scalability of neurosymbolic models presents a concern. Although these models have demonstrated promising outcomes in smaller-scale applications, scaling them to manage larger and more intricate tasks poses a substantial challenge \cite{Sarker2021NeurosymbolicAI, Onchis2022NeuroSymbolicClassifierSecurity}.
    
    NSAI holds the potential to transform NIDS by adaptively and dynamically countering new threats. Through continual updates to both neural and symbolic components, based on fresh data and interactions, NSAI systems could maintain a step ahead of cyber adversaries. Furthermore, the natural interpretability of symbolic reasoning provides cybersecurity professionals with insights into the rationale behind NIDS alerts, promoting faster and more efficient threat mitigation.

    In Section~\ref{sec:NSAITech}, we will provide an overview of NSAI methods. The method details can equip cyber researchers and practitioners with the necessary knowledge to effectively employ these advanced techniques in the context of NIDS. By thoroughly grasping the mechanisms and principles of NSAI, cyber professionals can tailor these strategies to better detect, analyze, and respond to the dynamic and increasingly sophisticated landscape of cyber threats.

\section{Neurosymbolic AI (NSAI) Techniques}
\label{sec:NSAITech}

Neurosymbolic AI, a fusion of symbolic and neural systems, has emerged as a promising approach to artificial intelligence \cite{Sarker2021NeurosymbolicAI, Kishor2022NeuroSymbolicAB}. This approach seeks to combine the strengths of both symbolic and neural systems, providing the interpretability and reasoning capabilities of symbolic systems with the learning and generalization capabilities of neural systems \cite{Susskind2021}. Neurosymbolic AI is a paradigm that integrates deep learning with traditional symbolic AI. These models originate from the need to decrease the quantity of data required for training by utilizing knowledge and reasoning concerning formerly gained and assimilated knowledge. The symbolic and neural approaches represent the two main branches of Artificial Intelligence (AI).
These two fields of AI have important differences. Neural approaches use multiple layers to progressively extract higher-level features from the raw input providing associative results by learning and adapting to the large amount of data provided; while symbolic ones produce logical conclusions and work best with relatively small and precise data, human intervention is in fact common in these methods.


The aim of NSAI is thus to reconcile the two main cognitive abilities, i.e., the ability to learn from the environment (neural part) and the ability to reason about what has been learned (symbolic part), thus combining the advantages of robust NN learning and the interpretability of symbolic reasoning \cite{garcez2019neural}. In NSAI knowledge is represented in symbolic form while learning and reasoning are processed by Neural Networks. Symbolic artificial intelligence refers to all methods that rely on high-level (human-readable) symbolic representations of problems, logic, and research. Reasoning is the process of generating conclusions from available knowledge using logical rules and principles. Knowledge can be represented in a variety of ways, most of which use variations of propositional and symbolic logic. Generally, knowledge is represented, either by a rule-based representation, that is, a set of IF-THEN rules are generated, or through First-Order Logic (FOL), then through predicates. Neural networks are nonlinear statistical structures utilized as modeling tools. They have the ability to simulate intricate relationships between inputs and outputs that cannot be represented by other analytical functions.

The integration of these two paradigms is an ongoing challenge; some of the main recent approaches will be outlined below. In order to simplify the collection of major SOTA works, we will define four areas: Relational Embeddings, End-to-end approaches, Hybrid integration NSAI, and Neural-Symbolic Reinforcement Learning. 

\subsection{Relational Embeddings}

In this subsection we will show approaches that map objects or concepts into continuous vector spaces, where distances or similarities between vectors reflect semantic or logical relationships between the objects themselves. These approaches use various machine learning techniques, such as neural networks, to learn such distributed representations of the data. 
These models seek to preserve semantic or logical relationships during the process of learning embedding vectors. They are models that embed logical or semantic relationships within the distributed representations of objects or concepts, enabling processing and inference on those relationships more efficiently and effectively. Embedding techniques are methods used in recent NSAI work \cite{bordes2011learning, sutskever2008using, santoro2017simple}. Training of these systems can occur through backpropagation. This has led research to focus on systems called \textit{relational embedding}, which are systems that represent relational predicates within neural networks. In this section we will focus on two approaches popular in recent years, LTN \cite{serafini2016logic} and NLM \cite{Valkov}, by way of explanation.

\subsubsection{Logic Tensor Network (LTN)}

Logic Tensor Networks represent a neurosymbolic formalism that integrates symbolic AI and neural computation, providing a unified language for various AI tasks \cite{serafini2016logic}. 
LTN introduces a differentiable version of first-order logic, termed Real Logic, into deep learning models. This integration allows for the use of logical operators in neural network computations, enabling the model to reason about the relationships between different inputs \cite{serafini2016logic}. 
One of the key advantages of LTN is their ability to combine symbolic and sub-symbolic representations within a single model. This capability allows for the incorporation of both structured and unstructured data within the same framework, proving particularly useful in domains such as natural language processing and computer vision \cite{badreddine2022logic}. LTNs extend Neural Tensor Network (NTN) \cite{socher2013reasoning}, a state-of-the-art relational embedding method. LTNs represent complex logical structure with a FOL formula. This approach adopts Real Logic as its formalism, an infinite-valued Fuzzy Logic language. Real Logic operates within a first-order language, $\mathcal{L}$, which includes $\mathcal{C}$, a set of constant symbols representing objects, $\mathcal{F}$, a set of functional symbols, $\mathcal{P}$, a set of relational symbols, and $\mathcal{X}$, a set of variable symbols.

\begin{figure*}[htbp]
    \centering
    \includegraphics[width=0.7\textwidth]{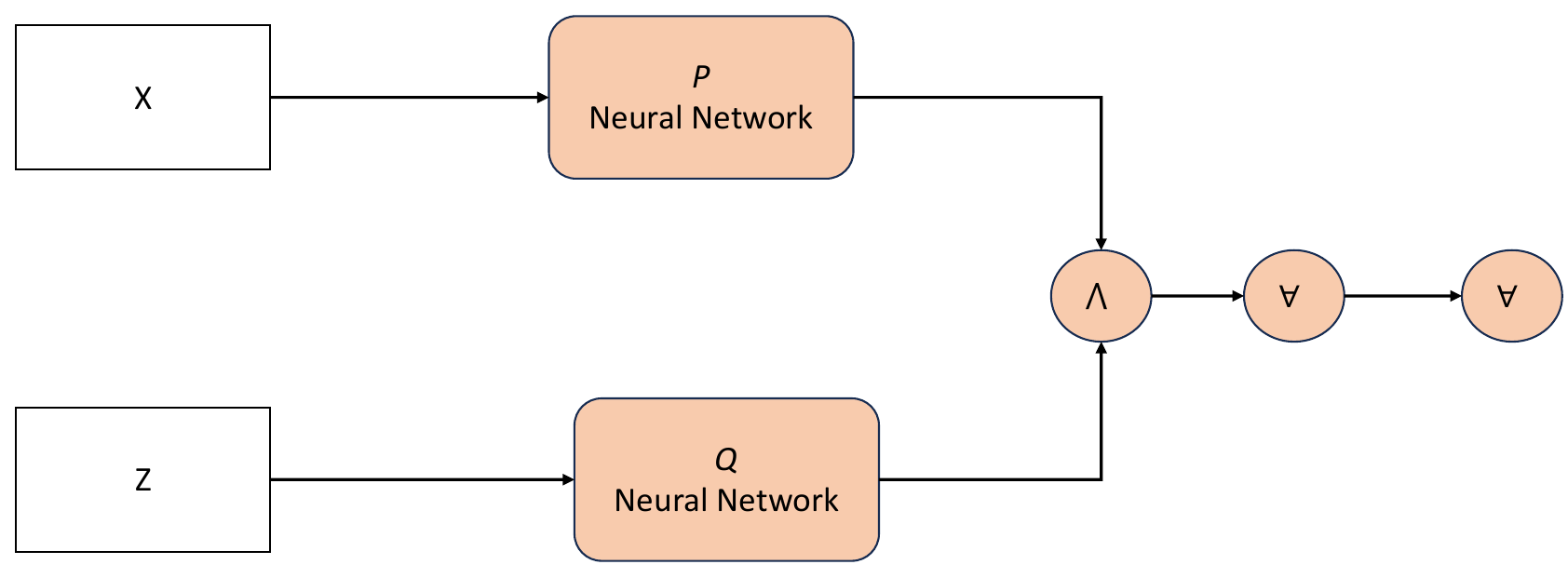}
    \caption{Logic tensor network for $\forall (x,z) (P(x)\land Q(z))$}
    \label{fig:LTN2}
\end{figure*}

In real-world scenarios there is often a degree of truth and exceptions are often present is interesting, formulas can take on intermediate values and a fuzzy semantics is used.
In this type of logic, domains are concretely interpreted by tensors in the real field. In logic, the term grounding refers to the operation of replacing the variables of a term or formula with constants or terms without variables; in \cite{garcez2019neural} it is also called \textit{instantiation}. With grounding, terms become tensors of real numbers. The formulas become real numbers restricted to the range $[0,1]$ (truth value).
With the use of real logic, objects can be represented by points in a feature space, and features and predicates are learnable. LTNs are trained to approximate the best satisfiability \cite{badreddine2022logic} making inference efficient with feedforward propagation. Fig \ref{fig:LTN2} shows an example of a LTN for $\forall (x,z) (P(x)\land Q(z))$.
In essence, LTN signifies an important advancement towards the development of more flexible and interpretable deep learning models, capable of human-like reasoning \cite{badreddine2022logic}.

\subsubsection{Neural Logic Machines (NLM)}
Neural Logic Machines (NLM) are a type of neurosymbolic model that combines the interpretability of symbolic reasoning with the power of neural networks. NLMs have been used in tasks such as sorting, path finding, and tower building, demonstrating the ability to learn and generalize from small amounts of data.
In this approach, through grounding, logical predicates are represented as tensors and can then be processed by neural networks. Grounding involves instantiating logical predicates with specific examples or input data. These examples are subsequently used to learn the parameters of the neural networks.
NLM generate conclusions or predictions based on the tensor representations of the logical predicates. This allows the symbolic reasoning capability of logical predicates to be combined with the powerful learning and generalization capabilities of neural networks \cite{Valkov}.

In this model, logical predicates are represented as probabilistic tensors. To achieve this representation, we consider a set of objects denoted as $\mathcal{U} = {u_1, u_2, \dots, u_m}$. A predicate $p(x_1, x_2, \dots, x_r)$ of \textit{arity} $r$ can be grounded on the object set $\mathcal{U}$, which is referred to as $\mathcal{U}$-grounding. This results in a tensor $p^{\mathcal{U}}$ with the shape $[m^r]\triangleq[m, m - 1, m - 2, \dots, m - r + 1]$. In this tensor, each entry $p^\mathcal{U}(u_{i1}, u_{i2}, \dots, u_{ir})$ indicates whether predicate p is true when its arguments equal to $u_{i1}, u_{i2}, \dots, u_{ir}$. To expand this representation to a group of predicates we use $C^{(r)}$ to denote the number of predicates with \textit{arity} $r$. We combine the grounding tensors of all predicates into a single $\mathcal{U}$-tensor with a shape of $[m^r, C^{(r)}] \triangleq [m, m - 1, m - 2, \dots, m - r + 1, C^{(r)}]$. The last dimension of the tensor corresponds to the predicates. Essentially, a set of unary predicates $C^{(1)}$ based on m objects can be described by a tensor with the shape of $[m, C^{(1)}]$, which represents a set of object properties. A tensor with a shape of $[m, m - 1, C^{(2)}]$ characterises a collection of {\it pairwise connections between objects} for binary predicates $C^{(2)}$. A maximum is imposed on the \textit{arity}, called the NLM's {\it breadth}. Furthermore, NLMs embrace a probabilistic perspective on predicates, with each entry in the basic tensor of $\mathcal{U}$ ranging from 0 to 1, indicating the probability of being true. 
NLMs use these tensors to represent premises, results and conclusions.

In summary, logical predicates are represented as probabilistic tensors through the $\mathcal{U}$-grounding process, where each entry represents the probability of the predicate being true. This representation enables modeling of object relations and performing probabilistic inferences using predicate tensors. The objective of NLMs involves constructing a neural architecture that learns rules capable of handling relational data with multiple arities while maintaining objectivity. Boolean logic rules are implemented in this neural architecture as lifted rules that comprise Boolean operations (AND, OR, NOT) on a set of predicates and quantifiers, together with logical quantifiers ($\forall$ and $\exists$). These neural units are then combined to form NLMs, ensuring a comprehensible and logical structure while using precise word choices, balanced language, correct grammar, and conventional structure. NLMs receive input tensors of predicates (premises), perform calculations layer by layer, and produce output tensors as conclusions. With an increase in the number of layers, higher levels of abstraction can be achieved. Thus, forward propagation in NLMs can be viewed as a sequence of rule applications.

The use of NLMs has been demonstrated in several tasks, highlighting their ability to learn from limited data and make meaningful inferences. By leveraging the strengths of symbolic reasoning and neural network processing, NLMs offer a promising approach to tackling complex problems that require relational reasoning and learning from data.

\subsubsection{Challenges and opportunities}
Both LTN and NLM enhance the development of NIDS by integrating expert knowledge, enabling the identification of complex and sophisticated threats, as well as recognizing previously unseen threats. This leads to more robust NIDS. NLMs, in particular, offer greater flexibility in learning from diverse data sources and improve interpretability. Conversely, LTNs use tensors to represent data, providing a multidimensional and rich representation of relationships between entities, though with somewhat less interpretability compared to NLMs. LTNs still offer more interpretability than traditional neural networks, yet deciphering model decisions can remain intricate. Both systems encounter challenges such as integrating with existing NIDS, high computational costs, and the risk of overfitting.

\subsection{End-to-end Approach}
In this subsection we will show approaches we call end-to-end; these approaches describe a system or architecture that addresses an entire sequence of operations independently, from input data to final output. This category represents a complete workflow, without dependencies on external components.
Below we will illustrate two different approaches. The first integrate probabilistic approaches, ProbLog and stochastic logic program, to neural networks. The second type integrates neural networks with discrete approaches. Both with an end-to-end approach, that is, a unified system in both training and inference.

\subsubsection{Integrating Probabilistic Logical Reasoning and Deep Learning}

Among the neuro-symbolic models that have emerged in recent years we find an approach that combines the strengths of Probabilistic Logic Programming (PLP) and deep learning. These models, such as DeepProblog \cite{manhaeve2021neural} and DeepStochLog \cite{winters2022deepstochlog}, have demonstrated their versatility in various domains, including image interpretation, probabilistic programming, and reasoning about complex representations.

By integrating PLP with deep learning techniques, these neuro-symbolic models demonstrate the ability to learn from data, grasp complex relationships, and reason effectively about uncertain information. They allow modeling of probabilistic dependencies and representation of rich, structured knowledge, enabling more nuanced and flexible reasoning abilities. DeepProbLog uses probabilistic logic programming to integrate logic programming with deep learning. In DeepProblog, logic rules are extended with probabilistic annotations that allow uncertainty to be incorporated into models \cite{manhaeve2021neural}.
DeepProbLog combines the ProbLog language with neural networks; in particular, it is based on an neural annotated disjunctions. In \cite{riguzzi2023foundations}, a NAD is defined as follows:

\begin{theorem}
\label{def:nad}
\textbf{Neural Annotated Disjunction (NAD)}. A NAD is represented by the expression:

\begin{equation}
    nn(m_r, \textbf{I},O,\textbf{d}) :: r(\textbf{I},O).
\end{equation}

In this formula, $nn$ is a reserved predicate, $m_r$ denotes a neural network with $k$ inputs and $n$ outputs, specifying a probability distribution across $n$ classes. $\textbf{I}=I_1,\dots,I_k$ represents a sequence of input variables for the NN, $O$ stands for the output variable, $\textbf{d}=d_1,\dots,d_n$ is a sequence of ground terms (the classes of the NN), and $r$ is a predicate referred to as the \emph{neural predicate}.

A \emph{ground NAD} takes the following form:

\begin{equation}
\begin{array}{l}
nn(m_r,\textbf{i},d_1) :: r(\textbf{i},d_1)~;~\dots~;
nn(m_r,\textbf{i},d_n) :: r(\textbf{i},d_n).
\end{array}
\end{equation}

Here, $\textbf{i}=i_1,\dots,i_k$ is a sequence of ground terms (the input to the NN), and $d_1,\dots,d_n$ are ground terms (the classes of this NN).
\end{theorem}

In the context of a ground NAD, the term $nn(m_r,\textbf{i},d_j)$ can be interpreted as a function that provides the probability of class $d_j$ when the network $m_r$ is evaluated with input $\textbf{i}$.
Consider the problem of assessing whether or not a poker hand is a poker, using a card images. DeepProbLog can be applied to this problem.
You could define a predicate $game([C_0, C_1, C_2, C_3, C_4], Outcome)$, where $[C_0, C_1, C_2, C_3, C_4]$ represent the images of the cards, and $Outcome$ is an integer, $1$ if hand is a poker, $0$ otherwise. Instances of this predicate would be in the form of atoms such as $game([\digit{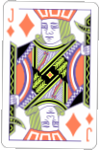}, \digit{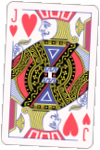}, \digit{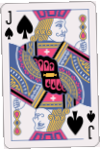}, \digit{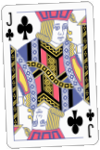}, \digit{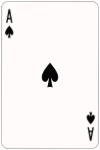}], 1)$.

For this example, we could have the NAD
\begin{equation}
    nn(m\_{card},[X],Y,[jack,queen,king,ace]) :: rank(X,Y).
\end{equation}
where the network that classifies single cards is represented by $m\_card$. 
To generate the grounding  we consider input image $\digit{jackQ.png}$ and obtain the ground NAD:
\begin{equation}
\begin{array}{l}
nn(m\_{card},[\digit{jackQ.png}],jack) :: rank(\digit{jackQ.png},jack)~;~\ldots~;~\\
nn(m\_{card},[\digit{jackQ.png}],ace) :: rank(\digit{jackQ.png},ace).
\end{array}
\end{equation}
Evaluating this would result in the instantiated NAD:
\begin{equation}
p_0 :: rank(\digit{jackQ.png},jack)~;~\ldots~;p_3 :: rank(\digit{jackQ.png},ace).
\end{equation}
where the output vector of the $m\_card$ network when evaluated on $\digit{jackQ.png}$ is $[p_0, \ldots, p_3]$.

A ground program is obtain from a DeepProbLog program by grounding probabilities and NADs. This ground program is then transformed into a ProbLog program. During inference, the program is initially grounded with respect to the query. Subsequently, ground NADs are generated from the NAD and are instantiated by conducting a forward pass on the neural network using the grounded input. This yields a ground ProbLog program. The program is then converted into a propositional formula, which is compiled and transformed into an arithmetic circuit. Finally, the circuit is evaluated to compute the probability.

In terms of learning in DeepProbLog, the objective is to determine the optimal values for both neural parameters and logic program parameters. This involves optimizing a loss function, such as mean square error, where the error measures the disparity between the probability assigned by the program to an atom and its labeled probability. Often, only atoms labeled with probability 1 are provided in DeepProbLog. In such cases, the aim is to maximize the probability assigned to these atoms, which can be achieved by minimizing the average of the negative logical probabilities of the atoms using the following loss function:

\begin{equation}
    \argmin_{\theta} \mathcal{L} = \argmin_{\theta}\frac{1}{|E|}\sum_{(q,1)\in E} -\log P_{\theta}(q)
\end{equation}

where $\mathcal{L}$ represents the loss, $\theta$ is the vector of all parameters, and $P_{\theta}(q)$ denotes the probability assigned by the DeepProbLog program to the query $q$. The learning problem is addressed using gradient descent, enabling the adjustment of both neural and probabilistic parameters in an integrated manner.

DeepStochLog \cite{winters2022deepstochlog} is another symbolic neural framework based on a type of stochastic logic programs, Stochastic Defined Clause Grammar (SDCG), which defines a probability distribution over possible derivations. The idea is the same as DeepProbLog, but integrated with SDCG rather than ProbLog. In DeepStochLog, similar to the $nn$ predicate of \cite{manhaeve2021neural}, the authors define a Neural Definite Clause Grammars (NDCG), i.e., an SDCG with the addition of the neural rule that, as in DeepProbLog, represents the NN.

These models have been used in tasks such as image interpretation and probabilistic programming, demonstrating the ability to learn complex representations and reason over them.

\subsubsection{NeurASP}

NeurASP is a model that combines logic programming and deep learning, specifically by integrating neural networks into the Answer Set Programming (ASP) framework \cite{bonatti2010answer}. Unlike DeepProblog and DeepStochLog, NeurASP uses deterministic rather than probabilistic inputs.

NeurASP introduces the concept of neural predicates, which are logical predicates extended with a neural component. These neural predicates allow neural networks to be used to perform inferences, learn from data, and enrich the symbolic reasoning process. NeurASP's neural networks generate input interpretations that are then used in the logic programming component. NeurASP attempts to combine the learning ability of NNs with the expressiveness of ASP for knowledge representation and reasoning. By allowing neural models to be incorporated into the ASP framework, it makes it possible to learn neural parameters from data and incorporate them into ASP applications.

The main concept of NeurASP is to create neural constraints using ASP rules, and then train neural models to satisfy these constraints. NeurASP can be used for a variety of tasks, including knowledge base completion, knowledge graph reasoning, natural language interpretation, and decision-making under uncertainty by merging logic-based reasoning with the learning capabilities of Neural Networks (NNs). The integration of neural networks into ASP allows complex data representations to be handled and patterns to be captured from large data sets, enriching the symbolic reasoning process \cite{yang2020neurasp}. Additionally, NeurASP can be used to train neural networks alongside rules so that they can learn from both explicit complex semantic constraints stated by ASP rules and implicit complex semantic correlations found in the data. Neural atoms can be used to feed the semantic loss \cite{Xu2018:NeurASP} from the reasoning module into neural networks. A neural network may learn more effectively as a result, even with less data. 

{\bf Answer-Set Programming:} A issue is encoded as a logic program in answer-set programming (ASP), a declarative problem-solving paradigm, so that the program's models, or answer sets, match to the problem's solutions. Dedicated ASP solvers can be used to compute answer sets. Here, give a brief overview of the key ASP concepts and make references to related literature \cite{Brewka2011:ASP}.

An ASP program is a set of rules of the form its 
\begin{equation}
\label{eq:rulesASP}
     a_1|\dots|a_k \leftarrow b_1,\dots,b_m,not \, c_1,\dots, not \, c_n
\end{equation}

A rule consists of a head and a body, where all $a_i, b_j, c_l$ are atoms, $not$ is negation and $k, m, n>=0$. The head of the rule is ${a_i,\dots,a_k}$, and the body conditions are ${b_1,...,b_m, not \ c_1,...,a_k}$. A rule states that if the atoms in the body are true, and there is no evidence of the negated atoms, then some atom in the head must be true. If $m=n=0$ and $k=1$, then the rule is a fact, and the right-hand side is omitted. 

A fact contains information that is always considered true, as its conditions are always met. A rule with an empty head is referred to as a constraint and is employed to filter out undesired answers.

An interpretation \(I\) is a set of ground atoms. It satisfies a rule \(r\) if it includes at least one \(a_i\) from the rule's head whenever it satisfies its body. Additionally, \(I\) is deemed a model of a program \(P\) if it satisfies to each rule in \(P\). An answer set of \(P\) is a model of \(P\) where each atomic element can be derived in a logically consistent and non-circular manner \cite{Gelfond1991:ASP}.

A program may have none, one, or multiple answer sets. For example, the simple program \[a \leftarrow\ \, not \, b. \, b \leftarrow\ \, not \, a.\] has the answer sets \(I_1 = \{a\}\) and \(I_2 = \{b\}\). If we introduce the constraint \(\leftarrow a\) to the aforementioned program, the sole remaining answer set is \(\{b\}\), as \(\{a\}\) is excluded.

{\bf Syntax of NeurASP:} 

NeurASP rapresents a neural network $M$ as a neural atom:
 
\begin{equation}
     nn(m(e,t), [v_1,...,v_n])
\end{equation}
here, a neural atom is denoted by $nn$, and the symbolic name of the NN $M$ is identified by $m$. A "point" of an input tensor to $M$ is represented by a list of terms $t$. Furthermore, $v_1,...,v_n$ represent all possible outcomes of each of $e$ random events.
A NeurASP program consists of ASP rules and neural atoms. 

{\bf Semantics of NeurASP}

In NeurASP, the answer sets ASP program and their probability is defined by the semantics from the NN ouputs.
To obtain ${\prod}'$ from each NeurASP program ${\prod}$, we replace each neural atom \ref{eq:rulesASP} with the set of choice rules
\[\{{m_i}(t)={v_1};...;{m_i}(t)={v_n}\}=1 \, for \, i{\in}\{1,...e\}.\]

By including neural networks, authors in \cite{Yang2020:NeurASP} provide a straightforward extension of answer set algorithms. The suggested NeurASP offers a straightforward and efficient method of integrating sub-symbolic and symbolic computation by treating the neural network output as the probability distribution over atomic facts in answer set programs, building on the idea of DeepProbLog.

\subsubsection{Challenges and opportunities}

End-to-end systems offer ways to handle uncertainty in data through the integration of logic and machine learning. These systems employ various strategies, such as probabilistic logic programming, stochastic logic programs, and answer set programming, to model and represent uncertainty, enabling them to address a wide range of real-world scenarios in cybersecurity. While they provide a certain degree of interpretability, deciphering model decisions can still be challenging.

These systems also offer greater flexibility in learning from heterogeneous data, allowing NIDS to manage diverse information sources. However, they present several challenges, including scalability and computational costs, as well as the complex integration with existing cybersecurity infrastructures. Successfully incorporating these approaches into a current NIDS framework necessitates close collaboration with experts in both cybersecurity and artificial intelligence.

In summary, end-to-end approaches manage uncertainty in data through the integration of logic and machine learning, employing various strategies to model and represent uncertainty. This capability allows them to effectively address typical real-world situations, including the intricate domain of NIDS.

\subsection{Hybrid Integration NSAI}

By hybrid integration, we refer to the approach of combining different components or approaches within a system or model. This may involve the use of machine learning models in combination with logical rules, or the incorporation of constraints into neural networks. These are hybrid approaches in which the two systems (neural and symbolic) are distinct, but whose results are combined and adapted to achieve better performance.  This type of NSAI model has two sub-models, one symbolic and one connectionist, so there is not a complete integration as in the other areas, but a collaboration between the two systems, as in Fadja et al. \cite{fadja2022neural} work, where a CNN is combined with a random forest throught Hierarcical Probabilistic Logic Programming \cite{nguembang2021learning} to obtain a more interpretabile system for early-stage prediction of critical state of Covid-19 patients. There are several frameworks that perform hybrid integration \cite{DBLP,Susskind2021}. Below we will give a description of a few as examples.

\subsubsection{Neuro-Symbolic Concept Learner (NSCL)}

The Neuro-Symbolic Concept Learner (NSCL) is a model that merges symbolic and neural components for tasks like visual reasoning and unsupervised learning of words, sentences, and visual concepts. 
The NSCL model comprises an object-based depiction of scenes and the translation of sentences into executable symbolic programs.
They use neurosymbolic reasoning to execute these programs on the latent scene representation \cite{DBLP}.

The CLEVR dataset served as the inspiration for the Neuro-Symbolic Concept Learner \cite{Johnson2016:clevrdataset}. The "image reasoning" dataset CLEVR contains images and a series of questions that go along with them. The model's output is the answers to these questions. There are cubes, cylinders, and spheres of various sizes, hues, and materials among the image samples in CLEVR.  Three submodels of NSCL are present, and each model is discussed below. 

{\bf Visual perception:}
The goal of the image parser is to produce "masks" of objects, or regions that are accurate to the pixel and have annotated colors, forms, and materials. For all objects, NSCL uses a pretrained Mask R-CNN \cite{He_2017:NSCL} to create object recommendations. The bounding box for every individual object is then passed along with the original picture to a ResNet-34 \cite{jian2016:NSCL} to separate the region-based and image-based features. To represent each entity, they are concatenated. In this scenario, by incorporating context information we can infer fundamental physical attributes of objects. This is possible by representing the entire scene

{\bf Concept quantization:} 
Identifying an object's qualities, such as its color or shape, is necessary for visual reasoning. The authors suppose that each visual characteristic, such as form, contains a collection of visual conceptions, such as Cube. NS-CL uses neural operators to implement these visual attributes by mapping the object's representation to an attribute-specific space. This procedure also requires the learning of concept vectors. The shape of the object is determined by calculating the cosine distances between these vectors. Similar to how we categorize concepts of relationships (e.g., Left) between two items, we concatenate the visual representations of the two objects to reflect the relationship between them.

{\bf DSL and semantic parsing:} 
An executable program with a hierarchy of primitive operations is generated from natural language queries by the semantic parsing module. This program is represented in a domain-specific language (DSL) created for visual question answering (VQA). Several functions for visual reasoning are included in the DSL. The operations can be composedly joined to create programs of unlimited complexity since they have a common input and output interface. 

{\bf Quasi-symbolic program execution:} The symbolic program executor generates predictions from tokenized questions and retrieved picture attributes. Formally, this model is "quasisymbolic" since the executor approximates non-differentiable functions probabilistically rather than logically as in a true symbolic AI model, which applies boolean and logical operations on data. The NSCL produces a vector as its output, and its $i$th member, $Mask_i \in [0,1]$, represents the likelihood that the matching object from the scene will appear in the output set. There are two benefits to using this representation. First, it permits several correct responses, which makes the questions the model is able to answer more complicated. Second, it entirely reverses the symbolic action, allowing for back-propagation. This method gets around genuine boolean circuits' constraints, which are typically irreversible. Due to this reversibility, concepts can be learned without explicit labeling by using concept embedding vectors, which are representations of object features.

\subsubsection{Neuro-Symbolic Dynamic Reasoning (NS-DR)}

The Neuro-Symbolic Dynamic Reasoning (NS-DR) model employs multiple autonomous submodels for feature extraction before feeding them into a final symbolic submodel. This model has shown promise in tasks such as dynamic reasoning and semantic parsing of sentences \cite{Susskind2021}. NS-DR is applied to the CLEVRER video reasoning dataset. CLEVRER improves on CLEVR by substituting videos for photos. NS-DR comprises a parser for video frames, a predictor for neural dynamics, a parser for questions, and an executor for programs \cite{Kexin2020:NSDR}. Details of the model are described in below.  

{\bf Video frame parser:} PropNet \cite{Wu2-19:PropNet} is a learning physics engine that can describe intricate item collisions. PropNet builds on earlier work in the field by simulating force propagation through numerous objects accurately and functions properly in the presence of incomplete information. Since all movies in CLEVRER are taken from a fixed camera position and objects are allowed to enter and exit the scene, functional accuracy with incomplete information is essential. For the NS-DR model, dynamics prediction offers the locations, trajectories, and collisions between objects.

{\bf Neural dynamic predictor:} The authors of \cite{Kexin2020:NSDR} used the Propagation Network (PropNet), a learnable physics engine that extends from interaction networks and updates a dynamic scene in an object- and relation-centric manner, for dynamics modeling. In order to forecast motion traces and collision occurrences, the model takes as input the object suggestions from the video parser and learns the dynamics of the objects over the frames. 

{\bf Question parser:} Unlike NSCL, the NS-DR model employs a newer Neural Machine Translation (NMT) model called Seq2Seq \cite{bahdanau2016:NSDR}. This model has displayed better accuracy with lengthy inputs. The model shares similarities with the question parser and uses a bidirectional LSTM encoder and an LSTM decoder with attention. 

{\bf Symbolic Program Executor:} Unlike the NSCL, the program executor of the NS-DR makes predictions using non-differentiable operations, making it a real symbolic model. Since back-propagation of error is impossible with non-differentiable operations, this model is unable to learn concept embeddings; instead, the frame parser must undergo supervised training to acquire concepts directly. This model operates internally much like a programming language interpreter, utilizing tokens to apply filter and reduction operations to the extracted video features. It is fully fixed-function and without any learned components.

\subsubsection{Challenges and opportunities}
Hybrid NSAI systems allow for the integration of expert knowledge from both neural and symbolic approaches, significantly enhancing NIDS. This integration combines the neural networks' capability to detect complex patterns with the clarity and interpretability of logical rules. Thanks to the symbolic component, these systems can perform logical reasoning on complex network data, identifying relationships and patterns that might be challenging to capture with solely neural methods. Additionally, hybrid NSAI systems are adaptable and can be easily updated to handle new threats or scenarios in the cybersecurity field.

However, the integration of neural and symbolic components makes the system complex to implement and manage, requiring additional skills and resources. Training such systems for NIDS may necessitate a well-balanced and representative dataset, as well as a complex optimization process to ensure the proper integration of neural and symbolic components. Furthermore, while there is an increase in interpretability compared to purely neural networks, interpreting the decisions of hybrid NSAI systems can still be complex due to the combination of neural and symbolic approaches.

Hybrid NSAI systems offer a promising combination of neural and symbolic approaches for NIDS, leveraging the strengths of both to enhance cybersecurity. However, it is crucial to consider the challenges associated with implementing, interpreting decisions, and training such systems before deploying them in real-world contexts.

\subsection{Neural-Symbolic Reinforcement Learning}
Neural-Symbolic Reinforcement Learning is an approach that combines the strengths of reinforcement learning and symbolic reasoning. This approach has been used in tasks such as game playing and robot navigation, demonstrating the ability to learn complex behaviors and reason over them \cite{kimura2021neuro}.

Neurosymbolic models and methods are also being developed to overcome critical challenges, such as the computational challenge of continuous validation of neural networks within a learning loop, a key concern for guaranteed safe deep reinforcement learning \cite{Anderson}. One promising solution is the "Estimate and Replace" strategy, which embeds calls to existing applications within deep learning architectures, as demonstrated by the EstiNet model's success with a table-based question-answering task, proving the method's ability to learn from less data \cite{hadash2018estimate}.

The fusion of deep learning with program synthesis to learn functions from data is gaining momentum in the arena of neurosymbolic programming \cite{Devlin, Chenxi2022}. Techniques such as "neural program synthesis" and "neural program induction" are leading the charge in crafting programs using neural networks conditioned on input/output examples or latent program representations \cite{sun2022neuro}.

Deep reinforcement learning is a prominent area in artificial intelligence, showing successful applications across various domains. Nevertheless, existing deep learning methods for reinforcement exhibit limitations in their ability to reason effectively \cite{yu2023survey}.

To tackle this issue, researchers have embarked on integrating symbolic knowledge into reinforcement learning approaches. Garnelo et al. \cite{garnelo2016towards} introduced a novel approach called Deep Symbolic Reinforcement Learning (DSRL) to address this challenge.

DSRL combines a neural back-end and a symbolic front-end, with the former processing raw sensor data into a symbolic representation, enabling the latter to devise effective strategies.

The DSRL framework comprises three essential components: a planner, utilizing symbolic knowledge to plan long-term actions (subtasks) for achieving intrinsic goals; a controller, employing DRL techniques to learn subpolicies for each subtask based on intrinsic rewards; and a metacontroller, which gains extrinsic rewards by assessing controller performance and suggesting new intrinsic goals to the planner.

Expanding on DSRL, Garcez et al. \cite{garcez2018towards} introduced Symbolic Reinforcement Learning with Common Sense (SRL+CS). This method enhances both learning and decision phases by modifying the reward distribution to account for the agent's interaction with the environment during learning. In the decision phase, the model assigns weights to individual Q functions based on the agent-environment distance, allowing for a comprehensive aggregation of Q values.

\subsubsection{Challenges and opportunities}
The integration of RL and symbolic logic enables the combination of RL's machine learning capability with the clarity and precision of expert logical rules. This approach allows NIDS to leverage the knowledge and expertise of cybersecurity experts alongside the adaptability and learning capabilities of neural networks. RL is particularly effective in making sequential decisions in dynamic and complex environments, such as those found in cybersecurity, where response actions depend on previous actions and the evolution of threats over time. The RL approach enables NIDS to adapt and learn from new threats and emerging scenarios, continuously improving its defense and response strategies.

However, integrating RL with symbolic techniques for NIDS can be extremely complex and require multidisciplinary skills, along with a significant and representative dataset. Additionally, a complex optimization process is needed to maximize the system's performance and ensure appropriate responses to threats. An NSAI system that integrates RL and symbolic logic offers significant advantages in managing complex cybersecurity threats but presents challenges in terms of implementation, decision interpretation, and optimal training. Nonetheless, if implemented correctly, it can be a powerful tool for enhancing NIDS capabilities, leading to more robust cybersecurity and data protection.

\section{Conclusion}
This article surveyed the current literature on NIDS, analyzing the ongoing efforts, results, and challenges in the field. Additionally, it has highlighted the potential use of NSAI, hypothesizing that these techniques, despite various challenges, could lead to improvements in this domain similar to those achieved in other fields.

We have specifically examined NSAI techniques and their state-of-the-art applications to assess their potential uses in NIDS and the associated challenges. NSAI techniques can enhance the explainability and robustness of NIDS by integrating domain knowledge and logic. However, they also pose several challenges, such as integration into existing systems, scalability, and the actual improvement of explainability.

Despite these hurdles, we contend that the adoption of NSAI techniques holds promise for significant advancements in NIDS. Specifically, NSAI could lead to better detection of unknown attacks and improved explainability, both of which are vital for the effectiveness and reliability of NIDS. This research suggests that further investigation into NSAI's integration within NIDS is not only warranted but could be pivotal in advancing the field.

\section*{Acknowledgments}
This work was supported in part by the U.S. Military Academy (USMA) under Cooperative Agreement No. W911NF-23-2-0108, the U.S. Army Combat Capabilities Development Command (DEVCOM) Army Research Laboratory under Support Agreement No. USMA 21050, and the Defense Advanced Research Projects Agency (DARPA) under Support Agreement No. USMA 23004. The views and conclusions expressed in this paper are those of the authors and do not reflect the official policy or position of the U.S. Military Academy, U.S. Army, U.S. Department of Defense, or U.S. Government.

\bibliographystyle{unsrt}  
\bibliography{references}

\end{document}